\documentclass[pre,showpacs,twocolumn]{revtex4}
\usepackage{amsmath,amsfonts,amssymb,wasysym}
\usepackage{bm}
\usepackage[dvipdfm]{graphicx}
\usepackage{color}

\newcommand{\etal}{ {\it et al}. }

\newcommand{\cF}{ {\cal F} }
\newcommand{\cV}{ {\cal V} }
\newcommand{\cK}{ {\cal K} }

\begin{document}
\title{
DNA as a one-dimensional chiral material. II. 
Dynamics of the structural transition between B form and Z form
}
\author{Teruaki Okushima}
\email{okushima@ike-dyn.ritsumei.ac.jp}
\affiliation{Research Organization of Science \& Engineering, 
Ritsumeikan University, Noji-higashi 1-1-1, Kusatsu 525-8577, Japan}
\author{Hiroshi Kuratsuji}
\affiliation{Research Organization of Science \& Engineering, 
Ritsumeikan University, Noji-higashi 1-1-1, Kusatsu 525-8577, Japan}

\date{\today}
\begin{abstract} 
We analyze the dynamics of
structural transitions between
normal right-handed B form
and 
unusual left-handed Z form
for a linear DNA molecule.
The dynamics under the external torque
in physiological buffer
is modeled by a Langevin equation,
with
the potential term 
given by the authors previously [Phys. Rev. E {\bf84}, 021926 (2011)].
With this model,
we first simulate the relaxation processes around B-form structure after
sudden changes of the external torques,
where
slow relaxation $\sim t^{-1/2}$
as a function of the elapsed time $t$ is observed.
Then,
the dynamics of structural transition from Z form to B form
is computed under various external torque strength.
For small external torques,
the transition proceeds via nucleation and the growth,
while 
for higher torques,
Z-form structure becomes unstable,
and the transition mechanism
is switched to a spinodal-like process.
These numerical results
are qualitatively
understood by simple phenomenological arguments.
\end{abstract}

\maketitle
\section{introduction}
Recent advances in experimental techniques 
shed light on 
the response of single double-stranded DNA molecules
to mechanical stresses, such as twisting and stretching
\cite{tenYears,structureTransition,Strick}.
Various theoretical models have been developed
to describe the mechanical responses of the molecule.
For example,
an extended theory of the classic elastic rod model 
can describe the supercoiling and its statistical properties
\cite{Strick,recentReviewMarko,MarkoSiggia,rodPlectoneme,supercoilDenaturation,twistStretch}.
Statistical models that phenomenologically describe various structural transitions 
were developed in Ref.~\cite{statModel}. 
Several mesoscopic models have also been developed, 
which can describe the interaction between DNA
conformation  and {\em melting} structural transition \cite{yan_marko,palmeri}.
In our previous paper \cite{okura}, guided by a gauge principle,
we have developed a mesoscopic model of DNA that describes 
the interplay between the global configuration and the various intrinsic
structures of DNA base pairs, such as B and Z structures.
In Ref.~\cite{okura},
we have also constructed 
an effective potential that describes B-Z transition of 
GC base-pair repeats,
and elucidated the statistical properties
that
the GC repeats
tend to make structural transition between
the usual right-handed B form 
and unusual left-handed Z form under small external torques,
in agreement with the recent experiment 
of Lee \etal~\cite{BZtrans}.
The GC repeats are frequently
included in 
DNA sequences of promoter regions,
and furthermore, 
in recent studies  \cite{z_cancer},
the Z-form structure is related to genetic instability.

In this paper we are concerned with the B-Z transition dynamics.
To construct the  dynamical model,
we first consider the Lagrangian for the linear DNA molecule
by introducing kinetic energy, in addition to the effective potential energy
given in Ref.~\cite{okura}.
From this Lagrangian,
the Euler-Lagrange equations of motions
are derived.
To model the effect of surrounding physiological buffer,
we moreover include viscous torques and random forces 
into the equations of motions.
The resulting equations of motion are utilized to 
analyze 
both
the relaxation dynamics 
and 
the B-Z transition dynamics
under various external torques.
An alternative approach to B-Z transition 
was proposed
in Ref.~\cite{phi4},
where
the structural transition dynamics
for an idealized dissipationless DNA molecule
were mediated by uniformly moving kinks.
(See the Appendix for the application of
the alternative treatment to our model.)
However,
as we see below,
the viscosity of the surrounding  solution
completely alters the transition mechanism.
In particular,
the transition mechanism
is switched 
from a statistically homogeneous nucleation process
to a spatially inhomogeneous spinodal-like 
dynamically unstable process,
as 
the external torque
is increased.

\section{Formulation\label{sec:model}} 
In the following,
we introduce 
a nonlinear dynamical model
that 
is a generalization of
the static model given in Ref.~\cite{okura}.
To this end,
we first construct the Lagrangian of a {\em linear} DNA molecule.
The total potential energy is 
a functional of
torsional angle of sugar-phosphate backbones $\chi(s)$
and
structural order parameter $\rho(s)$,
where 
$s$ 
is
the continuous parameter representing the number of bases from one end of the
molecule.
The potential energy density is
given by
\begin{equation}
\cF=\frac{C}{2}\left(\frac{d\chi}{ds}- \rho \right)^2+
\frac{d_1}{2}\big(\frac{d\rho}{ds}\big)^2
+\cV(\rho)
\label{eq:pot}
\end{equation}
where
$C=91 \times 10^{-20}$J,
$d_1=D_1/\omega_0^2$ with $D_1=4.1 \times 10^{-21}$ J and 
$\omega_0=0.6$ rad/bp
\cite{okura}.
The potential density $\cV(\rho)$
is given by
\begin{equation}
\cV(\rho) = \cV_0[(\rho/\omega_0)^2 - 1]^2+\tau_c {\rho},
\end{equation}
with
$\tau_c=-7.9 \times 10^{-21}$Nm \cite{marko2007} 
and
$\cV_0 = 6  \times 10^{-20}$J,
where the B-DNA and Z-DNA conformations
correspond to the minima at around
$\rho=\omega_0$ and
$\rho=-\omega_0$, respectively.
These parameter values are given in Ref.~\cite{okura} 
to reproduce thermal equilibrium properties.

The structural order parameter $\rho(s)$
is related to the
base-pair torsion,
namely,
the angle $\Theta(s)$ subtended at 
the center axis of linear DNA, as
\[
\frac{ d \Theta(s)}{ds} =\rho(s).
\]

Now besides the potential energy (\ref{eq:pot}), 
we shall take account of the kinetic energy terms coming from 
the angular variables $\Theta$ and $ \chi $.
These are introduced in the following manner:  
The first is concerning $ \Theta $.
The inertia moment
of a base pair $I_1$
is 
estimated 
as a rigid rod of length $L=2$nm ($\simeq$ the diameter of DNA),
approximately given by
\[
 I_1=\frac{M_1 L^2}{12}=1.47 \times 10^{-43}\  \text{kg m}^2,
\]
where $M_1$ is the total mass of a GC base pair, 
$262.228 $ g/mol $=4.4\times 10^{-25}$ kg,
which is estimated as 
the sum of the masses of bases G and C
($151.126$ and
$111.102$ g/mol,
respectively).
Using these,
the kinetic energy of base pairs
is 
given by
\[
 {\cal K}_1=\frac{I_1}{2} 
\left(
\frac{d\Theta}{dt}\right)^2.
\]

The second is concerning $ \chi $.
The inertia of moment
for 
sugar-phosphate backbone chain $I_2$
is similarly given by
\[
 I_2= M_2\times(1 \mathrm{nm})^2 = 7.8 \times 10^{-43}\ \text{kg m}^2,
\]
where
$M_2$ is the mass of backbone chain  per bp
given by $M_2=3.9 \times 10^{-25}$ kg,
which is
estimated 
as the sum
of two deoxyriboses
($2 \times$134.13 g/mol)
and
two
phosphoric acids
($2\times$98 g/mol).
Then,
the kinetic energy of sugar-phosphate backbone
per bp
is given by
\[
 {\cal K}_2=\frac{I_2}{2} \left(
\frac{d\chi}{dt}\right)^2.
\]

In single molecule experiments,
one end of a DNA molecule is fixed to a surface
and thus, without loss of generosity, 
we impose the following condition at the end of $s=0$:
\[
 \Theta(0)=\chi(0)=0.
\]
Furthermore,
suppose that the external torque $\tau$ is exerted on the other end $s=N$, 
where $N$ is the total bp length of the DNA molecule.
Hence,
the following potential is  added to the Lagrangian:
\[
 V_\mathrm{ext}= -\tau \chi(N).
\]
Putting all terms together,
the total Lagrangian is given by
\begin{align}
  L&=
   \int_{s=0}^N
  \left(\cK_1+\cK_2-{\cal F}
\right) 
   ds 
-V_\mathrm{ext}\nonumber\\
&\equiv K_1+K_2+F-V_\text{ext}
\label{eq:lagrangian}
\end{align}

For the sake of the numerical implementation,
we discretize the argument $s$
of dynamical variables $\rho(s,t)$ and $\chi(s,t)$,
into $s=0,\ d s,\ 2ds,\dots,nds(=N)$.
Then, using $\Theta_0=0$, 
we have
\[
 \Theta_i(t)=\sum_{j=1}^i \rho_j(t) ds
\]
and
\[
\frac{d}{dt} \Theta_i(t)=\sum_{j=1}^i \dot{\rho}_j(t) ds.
\]
Accordingly,
$K_1=\int_0^n \cK_1 ds$ is given by
\[
 K_1=\frac{I_1}{2}ds^3 
\sum_{i=1}^n 
\left[
\sum_{j=1}^i \dot{\rho}_j(t) 
\right]^2.
\]
Similarly,
we obtain the following discretized expressions for
$K_2$ and $F$, respectively:
\begin{align}
 K_2 &= \frac{I_2}{2} ds\sum_{i=1}^n  \dot{\chi}_i^2(t),\\
 F&=\frac{C}{2}
ds \sum_{i=1}^n 
\left(
\frac{\chi_i-\chi_{i-1}}{ds}-\rho_i
\right)^2\nonumber\\
&+\frac{d_1}{2}
\sum_{i=1}^{n-1}
\frac{(\rho_{i+1}-\rho_i)^2}{ds}
+\sum_{i=1}^n ds \cV(\rho_i).
\end{align}

The Euler-Lagrange equation 
$
 \frac{d}{dt} \frac{\partial K_1}{\partial \dot{\rho}_i}=-\frac{\partial
F}{\partial \rho_i}
$
gives
the equation of motion for $\rho_j$:
\begin{equation}
 I_1 ds^3 \sum_{k=i}^n \sum_{j=1}^k \ddot{\rho}_j(t)=f_i,
\label{eq:rho_dyn}
\end{equation}
where  $f_i=-\partial F/\partial \rho_i$ is given by
\begin{align}
f_i 
 =&Cds \left(\frac{\chi_i-\chi_{i-1}}{ds}-\rho_i
\right)
-d_1 ds
\frac{2\rho_i-\rho_{i-1}-\rho_{i+1}
}{ds^2}\nonumber\\
&
-ds \cV^\prime(\rho_i).
\end{align}

Similarly, from the Euler-Lagrange equation for $\chi_i$,
one obtains  the following equation of motion for $\chi_i$:
\[
 I_2 ds \ddot{\chi}_i=g_i
\]
where $g_i=-\partial F/\partial \chi_i$ is given by
\begin{align}
 g_i=&-C ds \left[
\frac{\rho_{i+1}-\rho_i}{ds}
-\left(
\frac{\chi_{i+1}-\chi_i}{ds^2}
-
\frac{\chi_{i}-\chi_{i-1}}{ds^2}
\right)
\right]\nonumber\\
&+\tau \delta_{i,n}.
\end{align}

Now we model the solvent effect to the DNA molecule
by Langevin equation.
For simplicity,
here we neglect the hydrodynamic long-range interactions
between DNA segments,
which would be important
especially for curved, long DNA molecules.
The Reynolds number ($=$inertial force/viscous force) is so small in
this situation
that
the viscous forces and thermal random forces 
thermally equilibrate the DNA molecule.
Then,
the viscous force is estimated as follows.
When a segment of $s$-th bp
rotates at angular velocity  $\omega=d \chi/dt$,
the solvent sticks to the DNA surface and the 
specific speed of fluid flow is $v \sim\omega R$,
where $R$ is the radius of DNA.
Hence,
we obtain the velocity gradient $\sim \omega R/R=\omega$,
and
the corresponding shear stress $\sim \eta \omega  $,
where 
$\eta $ is the viscosity coefficient.
In the following, 
the values of water ($1$ mPa$\cdot$s)
is used for $\eta$.
The viscous forces applied to the DNA segment
is given by
$ \kappa A \eta \omega$,
where
$A$ is the area of the DNA side surface and 
a dimensionless coefficient $\kappa$ is introduced.
For simplicity,
$A$ is approximated by
$2 \pi R h$, i.e., 
the area of side surface of cylinder with 
the radius $R$ of DNA ($\sim$ 1 nm)
and 
the height $h$  of the 1bp DNA segment($\sim 0.34$ nm).
Then, the viscous torque is given by
\begin{align}
&\text{(viscos force)}\times R =2 \pi \kappa R^2 h \eta \omega\nonumber\\
&\sim 2.14 \kappa\times 10^{-30} \text{kg m$^2$/s}
\times \omega\equiv \kappa \alpha \omega. 
\end{align}
Hence,
the equation of motion for $\chi_i$ is modified as:
\begin{equation}
I_2 ds \frac{d^2 \chi_i}{dt^2}=g_i-\kappa\alpha ds \frac{d \chi}{dt}
+\xi_i(t).
\label{eq:chi}
\end{equation}
Here,
the random forces $\xi_i(t)$ are introduced 
as Gaussian white processes satisfying,
\begin{equation}
\langle \xi_i(t)\xi_j(0)\rangle=
2\kappa\alpha ds \times k_B T \delta(t) \delta_{ij},
\label{eq:noise}
\end{equation}
where $k_B$ is the Boltzmann constant
and $T$ is the temperature of the surrounding buffer.
Note that 
Eqs.~(\ref{eq:chi}) and (\ref{eq:noise}) guarantee 
the $i$-th segment's rotational energy $I_2 ds \dot{\chi}(t)^2/2$
to reach the thermal value at temperature $T$.
Moreover,
we postulate that
the buffer  does not affect the dynamics of $\rho$,
because, in B form or Z form,
 base pairs are located inside the DNA molecule and
 apart from
the buffer.

In the  limit of $ds \to 0$,
one obtains the following coupled
stochastic partial differential equations:
\begin{align}
 & I_1 \int_{s_2=s}^{n} \int_{s_1=0}^{s_2}
 \frac{\partial^2\rho(s_1,t)}{\partial t^2} ds_1 ds_2\nonumber\\
&
=C
\left(
\frac{\partial\chi(s,t)}{\partial s} 
-\rho(s,t)
\right)
-\cV^\prime(\rho(s,t))
+d_1\frac{\partial^2\rho(s,t)}{\partial s^2}, 
\end{align}
and
\begin{align}
I_2& 
 \frac{\partial^2\chi(s,t)}{\partial t^2} \nonumber\\
=&C
\left(
\frac{\partial^2\chi(s,t)}{\partial s^2} 
-\frac{\partial\rho(s,t)}{\partial s} 
\right)
-
\kappa \alpha
 \frac{\partial\chi(s,t)}{\partial t} 
+\tilde{\xi}(s,t)\nonumber\\
&+\tau \delta(s-N),
\end{align}
where 
$\tilde{\xi}(i\ ds,t )=\xi_i/ds$ 
is scaled random noise satisfying
\[
\langle \tilde{\xi}(s_1,t_1)\tilde{\xi}(s_2,t_2)\rangle=
2 \kappa\alpha k_B T \delta(s_1-s_2,t_1-t_2).
\]

Note here that,
in the $ds \to 0$ limit with $\kappa=0$ and $T=0$,
the resulting
coupled wave equations for $\Theta$ and $\chi$ are 
formally equivalent to the model of Volkov \cite{volkov,yakushevich}.
These equations have a kink solution, as a special solution,
whose detailed derivation is given in the Appendix.

\section{numerical method\label{sec:method}}
In this section,
we develop the numerical method for solving
(\ref{eq:rho_dyn})
and 
(\ref{eq:chi}).
In the following,
we set $ds=1$ for simplicity.

First,
dimensionless variables are introduced
by measuring energy in units of $\cV_0$
and 
time in units of 
$\sqrt{I_1/\cV_0}(=1.5625\times 10^{-12} \text{ s})$.
These equations read:
\begin{align}
&\sum_{k=i}^n \sum_{j=1}^k \frac{d^2\rho_j(t)}{dt^{\ast 2}}
=C^\ast \left({\chi_i-\chi_{i-1}}-\rho_i\right)\nonumber\\
&+d_1^\ast (
2\rho_i-\rho_{i-1}-\rho_{i+1})
-4[(\rho/\omega_0)^2 - 1] \rho/\omega_0+\tau^\ast_c,
\label{eq:rho_dimless}
\end{align}
and
\begin{align}
I_2^\ast\frac{d^2 \chi_i}{dt^{\ast 2}}
&=-C^\ast \left[
{\rho_{i+1}-\rho_i}
-\left(
{\chi_{i+1}-\chi_i}
-
{\chi_{i}-\chi_{i-1}}
\right)
\right]\nonumber\\
&-\kappa \alpha^\ast \frac{d \chi}{dt^\ast}
+\xi_i(t^\ast)+\tau^\ast_{i,n},
\label{lan_chi}
\end{align}
where the asterisks
denote 
dimensionless quantities:
$C^\ast=C/\cV_0$,
$d_1^\ast=d_1/\cV_0$,
$\tau_c^\ast=\tau_c/\cV_0$,
$\tau^\ast=\tau/\cV_0$,
$I_2^\ast=I_2/I_1$, and
$\alpha^\ast=\alpha/\sqrt{I_1\cV_0}$.
Random forces $\xi_i(t^\ast)$ satisfy
\[
\langle \xi_i(t^\ast_1)
 \xi_j(t^\ast_2)
\rangle =2 \kappa \alpha^\ast T^\ast \delta(t^\ast_1-t^\ast_2) \delta_{ij},
\]
where
the dimensionless temperature $T^\ast$ is introduced as
\begin{equation}
T^\ast=k_B T/\cV_0  (= 0.07 \text{ at }300\text{K}).
\label{roomtemp}
\end{equation}

In the computation of Eq.~(\ref{eq:rho_dimless}),
it must be solved
for 
$\ddot{\rho}_i=\partial^2 \rho_i/\partial t^{\ast2}$.
Denoting
the right-hand side of 
Eq.~(\ref{eq:rho_dimless}) by $f^\ast_i$,
we rewrite (\ref{eq:rho_dimless})
as
\[
\sum_{j=1}^i \ddot{\rho_j}+
\sum_{k=i+1}^n \sum_{j=1}^k \ddot{\rho_j}=f^\ast_i
\]
Hence,
\[
\sum_{j=1}^i \ddot{\rho_j}
=f^\ast_i-f^\ast_{i+1} \equiv \delta f^\ast_i.
\]
We again rewrite this equation as
\[
\ddot{\rho_i}+
 \sum_{j=1}^{i-1} \ddot{\rho_j}
=\delta f^\ast_i,
\]
and obtain the following result:
\begin{equation}
\ddot{\rho_i}
=\delta f^\ast_i-\delta f^\ast_{i-1} \equiv \delta^2 f^\ast_i.
\label{eq:rho_i}
\end{equation}

In our computation,
Eqs.~(\ref{lan_chi}) and (\ref{eq:rho_i})
are numerically integrated.
At every $dt$-time-step computation,
each random force in Eq.~(\ref{lan_chi}) is set 
as $\xi_i dt=R$,
where
$R$ is a random number sampled
from the Gaussian distribution with
mean zero and variance $\sqrt{{2 \kappa \alpha^\ast T^\ast}dt}$.

\section{Dynamics of a linear DNA molecule}

\subsection{Slow relaxation\label{sec:relax}}
Here,
we study the relaxation after a sudden change of external torque.
To begin,
we  estimate two specific time scales:
relaxation time and oscillation period.
The specific relaxation time 
is given by
\[
t_\text{relax} \equiv  \frac{I_2}{\kappa \alpha}=3.6\kappa^{-1}\times 10^{-13} \text{s}
=0.23 \kappa^{-1} \sqrt{I_1/\cV_0},
\]
while
the time scale of oscillation
is
\[
t_\text{osc}\equiv \sqrt{\frac{I_2}{C}}=9.258\times 10^{-13} \times 10^{-12} \text{s}
=0.6 \sqrt{I_1/\cV_0}.
\]
Then, at first,
one might expect that
the dynamics were classified into two: 
over damped dynamics if $\kappa>1/3.6\sim 0.27$;
and under damped dynamics if $\kappa<1/3.6\sim 0.27$.
However, as we shall see below,
the dynamics show more complicated, slow relaxations $\propto t^{-1/2}$,
where
the relaxation time $t_\text{relax}$ will turn out to be
the key factor for determining the prefactor of the slow relaxations.
Hereafter,
we use dimensionless quantities and omit, 
for the sake of notational simplicity,
the asterisks on the symbols.

\begin{figure}[t]
\includegraphics[width=8cm]{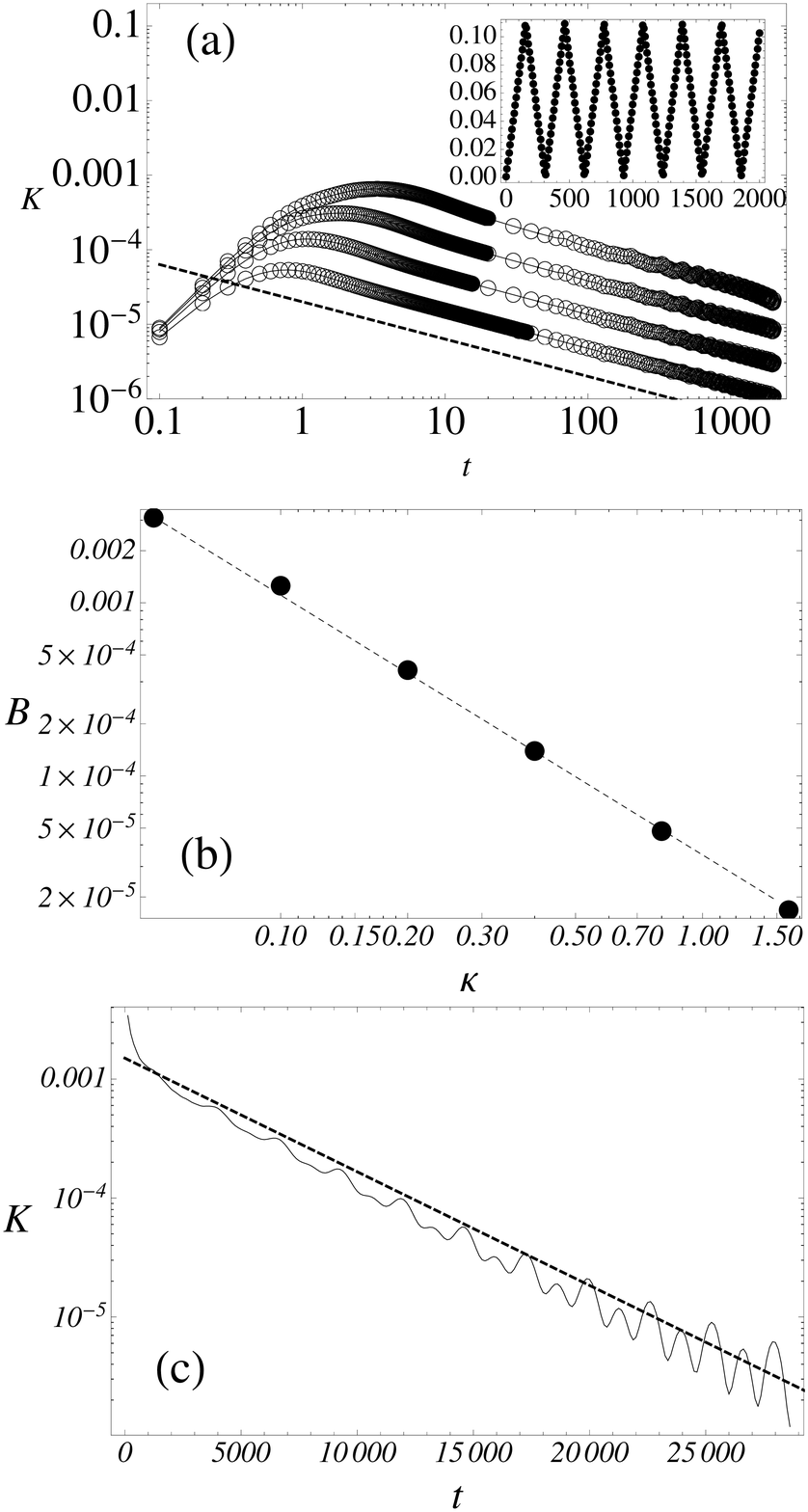}\\
\caption{
(a)
Total kinetic energies $K$ are plotted as functions of time $t$
after sudden addition of external torque $\tau=1$,
for $\kappa=0.1,0.2,0.4,0.8$ from top to bottom.
The dashed line $\propto t^{-0.5}$ is drawn for eye-guidance.
Long-time dynamics of $K$ for various $\kappa$ show slow relaxations $B(\kappa)t^{-1/2}$.
(Inset)
$K$ vs $t$ for the conserved system ($\kappa=0$).
(b)
Plot of coefficient $B$ as a function of $\kappa$.
The values of $B$ are obtained by fitting the data of $t>10$ in Fig.~(a)
to $B t^{-0.5}$.
The dashed line is $0.000035\kappa^{-3/2}$, which clearly shows  $B\sim
 \kappa^{-3/2}$.
(c) 
Semi-log plot of $K$ vs $t$   for 
the linearized system near $\rho=\omega_0$ with $\kappa=0.2$ and $\tau=1$.
The dashed line $\propto e^{-0.00022 t}$ shows the exponential decay.
}
\label{fig:relaxation}
\end{figure}
We simulated the relaxation dynamics 
at $T=0$.
The initial state was prepared as B-form structure:
namely,
\[
 \rho_i=\omega_0,\, \chi_i=\omega_0 i
\quad \text{for }i=1,2,\dots,n,
\] 
where $ds=1$ and the length of the DNA was set as $n=200$.
At $t>0$,
the external torque of $\tau=1$ 
was suddenly exerted on the $n$-th bp segment.
We computed
how
the DNA state relaxed to a  new stable state of
twisted B-form structure for various 
 dimensionless dissipation coefficients
$\kappa$.

Figure~\ref{fig:relaxation}(a)
shows 
the total kinetic energies $K=K_1+K_2$  as functions of time
for various $\kappa$.
For all $\kappa$ computed,
there are slow relaxations $B(\kappa)t^{-1/2}$, at their late times.
The prefactor $B(\kappa)$
is plotted in Fig.~\ref{fig:relaxation}(b),
as a function of $\kappa$.
We clearly see the relation $B(\kappa)\propto \kappa^{-3/2}$ holds.
This is qualitatively understood as follows:
As shown in the inset of Fig.~\ref{fig:relaxation}(a),
$K \propto t$ ($t<100$) for the conserved system of $\kappa=0$. 
Then, the nonconservative systems of $\kappa \neq 0$
would depart from the behavior of the conserved system
at the relaxation times $t_\text{relax}$.
At the times,
the kinetic energy $K$
would also
have their
maximum values
$K \propto t_\text{relax} \sim \kappa^{-1}$.
After the times, 
$K$ show slow relaxation $K \propto t^{-1/2}$.
Putting these estimates together,
 the following relation is  derived:
\[
 K\sim \kappa^{-1}\left(\frac{t}{t_\text{relax}}\right)^{-1/2}
\sim \kappa^{-3/2} t^{-1/2}.
\]
This 
qualitatively 
agrees with the above mentioned numerical result
$B(\kappa)\propto \kappa^{-3/2}$.

Notice that
the relaxations studied here
are not usual exponential
but an anomalously slow one $\sim t^{-1/2}$.
This type of slow relaxation is well known 
in the studies on one-dimensional nonlinear dynamical systems 
(e.g., Ref~\cite{slow1D}).
For comparison,
the torque response of the linearized equations near $\rho=\omega_0$
is plotted in Fig.~\ref{fig:relaxation}(c),
from which
we see that
$K$ relaxes exponentially with $t$.
Hence,
the origin of the observed power-law relaxation dynamics
is the nonlinear coupling in quasi-one-dimensional 
 DNA molecules.

\begin{figure}[t]
\includegraphics[width=6cm]{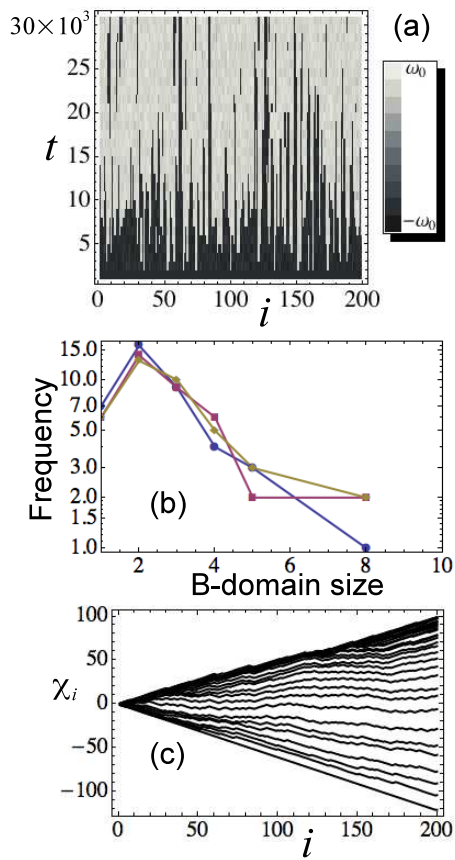}
\caption{
(Color online)
Structural transition dynamics from Z to B form,
for $\tau=0$ and $\kappa=0.01$.
(a)
Density plot of $\rho_i$
as functions of dimensionless time $t$ and $i$.
(b)
Semi-log plot of frequency distributions of B-domain sizes for 
$t=799000(\bullet)$, $839000(\blacksquare)$, $879000(\blacklozenge)$
(c)
Snapshot graphs of $\chi_i$ vs $i$
at dimensionless time $t=0,\ 10^5,2\times 10^5,\dots,20\times 10^5$
are plotted from below to top, respectively.
}
\label{fig:chinsnap}
\end{figure}

\subsection{Transition from Z form to B form\label{sec:zbtrans}}
Now we consider the dynamics of
structural transition from 
Z form to B form.

The initial state was prepared as Z-form structure:
namely,
\[
 \rho_i=-\omega_0,\, \chi_i=-\omega_0 i
\quad \text{for }i=1,2,\dots,n,
\] 
where the length of the DNA was set as $n=200$.
At $t>0$,
the external torque $\tau \geq 0$ 
was suddenly exerted on the $n$-th bp segment.
We computed
how
the DNA state relaxed to the stable B-form structure
from the initial Z-form structure,
at the room temperature $T=0.07$ [see Eq.~(\ref{roomtemp})].

The transition dynamics from Z form to B form 
is shown in 
Fig.~\ref{fig:chinsnap}(a)-\ref{fig:chinsnap}(c), 
for $\tau=0$ and $\kappa=0.01$.
From Fig.~\ref{fig:chinsnap}(a),
we clearly see the following:
(1)The stable B-form domains, indicated in light gray in the figure,  are quickly created 
from the beginning of time evolution.
(2)
Then,
the mixed states of B-form and Z-form structures are formed,
where
the statistical weights of B form grow
in the course of time evolution.
(3)
Eventually,
almost all segments become B-form structures.
Note here that,
if the nucleation is a random process with probability $p$,
the frequency of domain size $D$ should scale as $ \propto p^D$.
Figure~\ref{fig:chinsnap}(b)
shows that
this scaling relation indeed holds
for B-domains larger than two segments
and that
the critical size of the nucleation
is two segments.
As a result of the statistically uniform nucleation,
$\chi_i$ changes with keeping the linear shape $\chi_i \propto i$
[Fig.~\ref{fig:chinsnap}(c)].
More preciously, the following approximate relation holds:
\[
 \chi_i \simeq i \langle \rho \rangle +\text{fluctuating term},
\]
where $ \langle \rho \rangle $
denotes the average value of $\rho_i$.
This is easily understood
because
$\chi_i \simeq \sum_{j=1}^i \rho_j$
holds,
and 
because
one can replace
$\rho_j$ in this equation by $\langle \rho \rangle$
for the statistically uniform nucleations.

\begin{figure}[t]
\includegraphics[width=6cm]{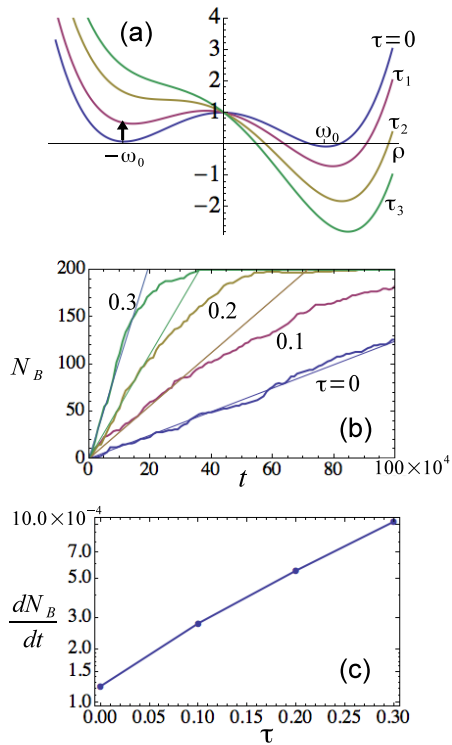}
\caption{(Color online)
(a)
Effective potentials for $\rho$ 
are plotted for various external torques $\tau=0,\ \tau_1,\ \tau_2,\ \tau_3$,
where $0<\tau_1<\tau_2=2.65<\tau_3$.
For $\tau=\tau_1$,
the activation energy required for transition from Z form to B form
is lowered by the amount indicated by the vertical arrow.
Larger external torque $\tau=\tau_3$
makes the Z-form structure  unstable.
$\tau_2$ is the critical value at zero temperature
beyond which 
the Z-form structure is no longer metastable
but
becomes unstable.
(b)
Number of B-form segments $N_B$ is plotted as a function of dimensionless time,
for $\tau=0,\ 0.1,\ 0.2,\ 0.3$.
Thin lines indicate the linear fitting to the data.
(c)
The speeds of $N_B$ increases
are semi-log plotted as a function of $\tau$.
}
\label{fig3}
\end{figure}

Now 
we consider  how  the nucleation rate depends on the external torque strength.
When a small torque is exerted on a terminal,
the interactions between neighboring segments
will quickly induce torque balances on all the segments.
Then,
the effective potential for structural parameter $\rho$
must be modified by the exerted torque,
as depicted in Fig.~\ref{fig3}(a).
Under  a weak external torque $\tau$ satisfying $0<\tau<\tau_2$,
the activation energy for transition from Z form to B form 
is given by
\[
 \Delta E \simeq \Delta E_0 -\Gamma \tau,
\]
where $\Delta E_0$ is the activation energy when $\tau=0$.
Accordingly,
nucleation rate $p$ 
must depend on $\tau$ as follows:
\begin{equation}
p \propto \exp(-\beta \Delta E)\propto 
\exp(\beta \Gamma \tau).
\label{eq:p}
\end{equation}
Figure~\ref{fig3}(b)
plots
the total numbers of B-form segments,
$N_B$,
as functions of time,
for various $\tau$.
This figure
shows that
each $N_B$ initially increases linearly with time,
which is
consistent
with the above confirmed fact that
the creations of B-form domains 
are mediated via statistically uniform nucleation mechanism
for weak external torques.
From the fitting lines in Fig.~\ref{fig3}(b),
we obtained the speeds of initial increases,
$d N_B/dt$.
Figure \ref{fig3}(c)
plots  the initial $d N_B/dt$ as a function of $\tau$.
This clearly shows
the exponential dependency of the nucleation rate to $\tau$,
as expressed in Eq.~(\ref{eq:p}).

\subsection{Transition under high external torque\label{sec:hightorque}}
When the external torque exceeds a critical value of 
$\tau_2 \simeq 2.65$, however,
Z-form structure
changes from metastable state 
to
unstable state,
as shown in Fig.~\ref{fig3}(a).
Hence,
for such  high external torque conditions,
the structural transition would
proceed via a mechanism other than the uniform nucleation.

\begin{figure}[h]
\includegraphics[width=5.2cm]{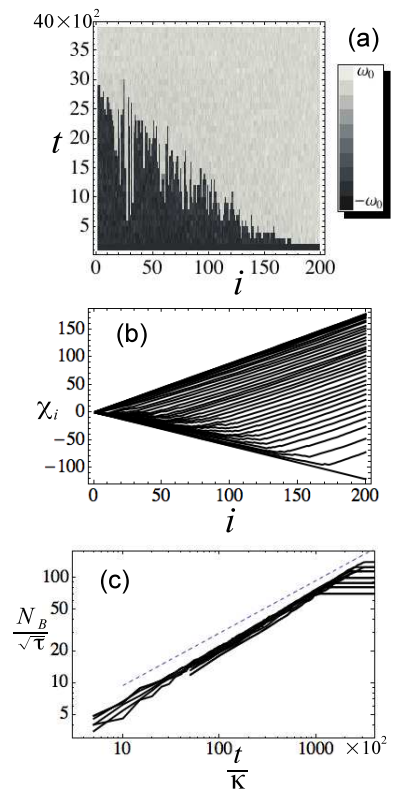}
\caption{
B-Z transition dynamics under high external torques.
(a)
Density plot of $\rho_i$ for $\tau=3$ and $\kappa=0.01$,
as functions of dimensionless time $t$ and $i$.
(b)$\chi_i$ vs $i$ for $\tau=3$ and $\kappa=0.01$.
The curves respectively correspond to $t=0,\ 100,200,\dots,2000$ results
from below.
(c)
$N_B/\sqrt{\tau}$ are plotted as functions of $t/\kappa$
for $(\tau,\kappa)$ in
$\tau=2.5,\ 3,\ 4,\ 5,\ 6,\ 8$
and $\kappa=0.01,\ 0.1,\ 1$.
All data are scaled onto a master curve.
The dashed line $\propto t^{0.5}$ is drawn for eye guidance
(see text).
}
\label{fig4}
\end{figure}

For the  high external torques,
the transition dynamics from Z-form to B-form are shown 
in Fig.~\ref{fig4}(a)-\ref{fig4}(c).
From Fig.~\ref{fig4}(a),
we see that the structural transitions occur about 100 times faster
than that for  the small external torque cases ($\kappa=0.01$).
In particular,
the transition is not uniform:
It first happens
at the terminal on which the external torque is exerted,
and then
the interface between Z form and B form
moves to the other terminal.
Due to the B-Z interface-mediated transition,
$\chi_i$ makes a kink at the interface, 
as shown in Fig.~\ref{fig4}(b).
Figure~\ref{fig4}(c)
shows
the number of B-domain segments,
$N_B$, as functions of time,
where all $N_B$ computed here
are fitted well
to the following form
\begin{equation}
\frac{N_B(t)}{\sqrt{\tau}} \propto \left(\frac{t}{\kappa}\right)^{0.5}.
\label{powernumeric}
\end{equation}

This is qualitatively 
understood from the way of B-form domain growth.
Suppose, at a time $t$,
$x$ base-pairs from the terminal
have already made transitions into a B-form structure.
At the moment,
the Z-form segment adjacent to the B-Z interface
is about to transition into the B form.
Let the transition be completed within a time interval $t_\text{trans}$.
Then,
the sweep speed $v$ of the B-Z interface
is given by $1/t_\text{trans}$.
In the following,
we will estimate $t_\text{trans}$.
At the beginning of this transition,
the angular velocity of the motion is assumed to be zero,
because of the large viscosity.
Moreover, the B-form DNA of length $x$ bp is approximated by
a rigid rod of inertia momentum $xI=x(I_1+I_2)$.
Then, its angular velocity $\Omega$ obeys the following equation
\[
xI \frac{d\Omega}{dt}= \tau-x \kappa\alpha  \Omega,
\]
resulting in
\[
 \Omega=\frac{\tau}{x\kappa\alpha}(1-e^{-\frac{\kappa\alpha}{I} t}).
\]
Solving the following equation for $t_\text{trans}$:
\[
2 \omega_0=\int_0^{t_\text{trans}} \Omega dt,
\]
one obtains 
\[
 t_\text{trans}=\sqrt{\frac{4 \omega_0 x}{\tau I}}+\frac{\kappa\alpha}{6} 
\left(
\frac{4 \omega_0 x}{\tau I}
\right)
+o[\kappa^2].
\]
With this expression,
the sweep speed is given by $v=1/t_\text{trans}$.
Integrating 
$dx/dt=v=1/t_\text{trans}$,
we obtain the following relation:
\[
 \sqrt{\frac{4 \omega_0}{\tau I}} x^{3/2}+
\frac{\kappa\alpha}{3} 
\frac{
 \omega_0 x^2}{\tau I}=t.
\]
For large $x$, we have
\[
 \frac{x}{\sqrt{\tau I}} \sim \left(\frac{t}{\kappa\alpha}\right)^{1/2},
\]
and therefore
Eq.~(\ref{powernumeric}) holds.

We have confirmed that
the unstable segment adjacent to the B-Z interface
induces the structural transition dynamics.
Hence, the transition dynamics can be called
a kind of real-space spinodal decomposition,
in contrast to the usual spionodal decomposition 
that occurs in reciprocal space.

\section{conclusion}
In this paper,
we studied  the dynamics of a linear DNA molecule.
To this end,
the statistical model given in  Ref.~\cite{okura}
is generalized by including rotational kinetic energies.
Then,
the dynamics under the external torque in a surrounding physiological
buffer is modeled by a Langevin equation.

In Sec.~\ref{sec:relax},
by using the Langevin dynamics,
we first simulated
the relaxation phenomena after 
sudden additions of external torques.
We found that
the total kinetic energy 
shows non-exponential slow relaxation 
as
$K=Bt^{-1/2}$,
which is a
characteristic of one-dimensional nonlinear dynamical systems.
In contrast,
$K$ relaxes exponentially with $t$ for the { linearized} equations of motions,
which means that
the nonlinearity is indispensable for the non-exponential decay.
The prefactor $B$ turned out to depend on 
viscosity proportional constant $\kappa$,
as $B\sim \kappa^{-3/2}$.
We found that
this can be qualitatively understood from the fact that
the dynamics of nonconservative systems depart from the behavior of 
the conserved system at the relaxation times of
$t_\text{relax}\sim 1/\kappa$.

We then proceeded to
study the dynamics of structural transition from Z form to B form
under external torques, $\tau$, exerted on a terminal [Sec.~\ref{sec:zbtrans}].
For small external torques,
B-form domains nucleate and grow.
This statistical process occurs uniformly within the whole DNA segments.
The speed of B-domain growth thus obeys an Arrhenius equation
with the activation energy $\Delta E=\Delta E_{0}-\Gamma \tau$,
where 
$\Delta E_0$
is the activation energy for $\tau=0$
and
$\Gamma$ is 
the linear response coefficient of the activation energy 
to the external torque $\tau$.

In contrast to this,
when the external torque is larger than a critical value,
Z-form structure becomes unstable rather than metastable.
For such a high external forces,
the transition mechanism is switched to 
a spatially inhomogeneous spinodal-like process (Sec.~\ref{sec:hightorque}). 
Namely,
the transition first occurs at the terminal which  the external torque acts on.
Then, the B-Z interface created there
starts to sweep to the other terminal.
When the interface reaches  the other terminal,
the whole transition is completed.
In this case,
the number of B-form segments
scales as $\propto t^{1/2}$ with time $t$.
We gave a phenomenological estimate
that accounts for
this numerically obtained scaling,
in which
we assumed that
the B-form DNA is approximated by a rigid rod
and
that
the transition into B form, 
occurring at the Z-form segment adjacent to the B-Z interface,
gives rise to the rotation of the B-form rod.
Using these two assumptions, we reproduced the numerically obtained
scaling relation as
$
 {x}/{\sqrt{\tau I}} \sim \left({t}/{\kappa\alpha}\right)^{1/2}
$.

We expect that
the simple dynamical model developed in this paper
equips the essence of interaction
between 
DNA chiral structure and the mechanical response to the external forces,
and hence we believe
that
the various structural transition phenomena elucidated in this paper
can be verified in future experiments.

It is noteworthy to remark that
we have neglected 
the hydrodynamic interaction between DNA segments
in modeling the effect of physiological buffer,
as discussed in Sec.~\ref{sec:model}.
This effect would become important
especially for studying 
dynamical interaction between
structures (e.g., B form, Z form)
and 
three-dimensional configurations (e.g., plectoneme) of a DNA molecule,
because
two segments, far apart from each other in a nucleic acid sequence,
can be spatially close to each other.
The inclusion of this effect is a further task of this work.

As a final remark,
we note that the dynamical aspects we have developed  
will provide a useful tool for investigating the dynamics of 
other structural transitions, such as
a denaturation process
\cite{PB,cocco1999,yakushevich}.

\appendix
\section{Kink propagation in a conserved system\label{sec:appendix}}
In this appendix,
we show
that
kink propagations
occur in the conserved system (i.e, $T=0,\kappa=0$).
As an ideal case,
we consider a DNA molecule of infinite length without external torques.
From the principle of least action,
we derive equations of motions for 
$ \Theta(s,t)$ and  $\chi(s,t) $.
The action is given by
$S = \int^{t}_0 L dt$,
where 
$L$ is
the Lagrangian of Eq.~(\ref{eq:lagrangian}).
The first variation of $S$
is given by
\begin{eqnarray}
\delta  S & = & \int\int \big[  \{-I_1 \frac{\partial^2\Theta}{\partial t^2} -C\frac{\partial^2}{\partial s^2}
(\chi - \Theta)\nonumber\\ 
    &-&d_1\frac{\partial^4\Theta}{\partial s^4} + \frac{\partial}{\partial s}\big(\frac{\partial \cV}{\partial \rho}\big)\} \delta \Theta 
    \nonumber \\
    & + & \{-I_2 \frac{\partial^2\chi}{\partial t^2} + C\frac{\partial^2}{\partial s^2}(\chi - \Theta)
    \}\delta\chi \big] dsdt.
\end{eqnarray}
Arbitrary $ \delta \chi$ and  $\delta\Theta $ should satisfy $\delta S=0$,
which gives the following equations of motion:
\begin{align}
&  -I_1 \frac{\partial^2\Theta}{\partial t^2} -C\frac{\partial^2}{\partial s^2}(\chi - \Theta) 
    -d_1\frac{\partial^4\Theta}{\partial s^4} + \frac{\partial}{\partial s}
    \big(\frac{\partial \cV}{\partial \rho}\big)  =  0  \label{eq:kink1} \\
&-I_2 \frac{\partial^2\chi}{\partial t^2} + C\frac{\partial^2}{\partial
 s^2}(\chi - \Theta) =  0
\label{eq:kink2}
\end{align} 
To obtain kink solutions with velocities $v$,
we set 
\[
 \chi(s, t) = F(s-vt),\ \Theta (s,t) = G(s-vt)
\]
in Eqs.~(\ref{eq:kink1}) and (\ref{eq:kink2}).
Thereby,
these equations reduce to 
the following ordinary differential equations
for $F(x)$ and $G(x)$ with $ x = s- vt $:
\begin{align}
&   -I_1v^2\frac{d^2G}{dx^2} - C\frac{d^2}{dx^2}\big(F-G) - d_1\frac{d^4 G}{dx^4} 
+ \frac{d}{dx}\big(\frac{d\cV}{d\rho}\big)  =   0, \\
&   -I_2v^2 \frac{d^2 F}{dx^2} + C\frac{d^2}{dx^2}(F - G)  =  0.
\end{align}
By eliminating $ F $ from these equations,
the equation for $G$ is given by
\[
d_1\frac{d^4G}{dx^4} + A\frac{d^2G}{dx^2} -\frac{d}{dx}\big(\frac{d\cV}{d\rho}\big) = 0,  
\]
with 
\begin{equation}
A = I_1v^2 + \frac{I_2v^2C}{C-I_2v^2}.
\label{eq:A}
\end{equation}
This equation gives a first integral
\[
    d_1\frac{d^3G}{dx^3} + A\frac{dG}{dx} - \frac{d\cV}{d\rho} = C_1.  
\]
By setting $  g(x)=\frac{dG(x)}{dx}  $,
one obtains
\[
 d_1\frac{d^2 g}{dx^2} + A g - \frac{\partial \cV}{\partial \rho}(g)  = C_1.
\]
This is the equation of motion for a particle of mass $d_1$ 
moving in a potential of $\frac{A}{2} g^2 -\cV(g)-C_1 g$.
Hence,
we obtain 
the following another first integral:
\begin{equation}
 C_2= 
 \frac{d_1}{2}\big(\frac{dg}{dx}\big)^2 + \frac{A}{2}g^2 - \cV(g) -C_1g.
\label{eq:kink0}
\end{equation}

According to 
Jensen \etal~\cite{phi4,saltNote},
here we assume the following symmetric potential
$
 \cV(\rho) = \cV_0\left[
\left(
{\rho}/{\omega_2}
\right)^2 - 1)
\right]^2.
$
For Eq.~(\ref{eq:kink0})
to allow for kink solutions,
we have to choose $C_1$ and $C_2$ as follows:
\[
C_1=0,\quad
 C_2=\frac{\omega_0^4}{\cV_0}
\left(
\frac{\cV_0}{\omega_0^2}+\frac{A}{4}
\right)^2-\cV_0.
\]
The resulting equation 
is given by
 \begin{equation}
\left(
\frac{dg}{dX}
\right)^2
= (g^2 - a^2)^2, 
 \end{equation}
where $a^2=\omega_0^2+\frac{\omega_0^4A}{2 \cV_0}$
and 
$X=x/\sqrt{d_1 \omega_0^4/2 \cV_0}$.
This equation has kink solutions 
$
   g(X) = \pm a\tanh(aX)
$.

Note here that
the family of kink solutions 
exist for arbitrary $v$ with $v^2<C/I_2$ [see Eq.~(\ref{eq:A})],
which is in contrast to 
the result of nonconservative systems,
where 
the speeds of interfaces
were uniquely determined with the viscosity
(Sec.~\ref{sec:hightorque}).

\end{document}